\documentclass[]{article}
\usepackage{lmodern}
\usepackage{amssymb,amsmath}
\usepackage{ifxetex,ifluatex}
\usepackage{fixltx2e} 
\ifnum 0\ifxetex 1\fi\ifluatex 1\fi=0 
  \usepackage[T1]{fontenc}
  \usepackage[utf8]{inputenc}
\else 
  \ifxetex
    \usepackage{mathspec}
  \else
    \usepackage{fontspec}
  \fi
  \defaultfontfeatures{Ligatures=TeX,Scale=MatchLowercase}
\fi
\IfFileExists{upquote.sty}{\usepackage{upquote}}{}
\IfFileExists{microtype.sty}{%
\usepackage{microtype}
\UseMicrotypeSet[protrusion]{basicmath} 
}{}
\usepackage[margin=1in]{geometry}
\usepackage{hyperref}
\hypersetup{unicode=true,
            pdftitle={Eliciting the Endowment Effect under Assigned Ownership},
            pdfauthor={Patrick Barranger, Rohit Nair, Rob Mulla, Shane Conner},
            pdfborder={0 0 0},
            breaklinks=true}
\usepackage{color}
\usepackage{fancyvrb}

\DefineVerbatimEnvironment{Highlighting}{Verbatim}{commandchars=\\\{\}}
\usepackage{framed}
\definecolor{shadecolor}{RGB}{248,248,248}
\newenvironment{Shaded}{\begin{snugshade}}{\end{snugshade}}

\newcommand{\CommentTok}[1]{\textcolor[rgb]{0.56,0.35,0.01}{\textit{#1}}}

\newcommand{\DataTypeTok}[1]{\textcolor[rgb]{0.13,0.29,0.53}{#1}}
\newcommand{\DecValTok}[1]{\textcolor[rgb]{0.00,0.00,0.81}{#1}}

\newcommand{\FloatTok}[1]{\textcolor[rgb]{0.00,0.00,0.81}{#1}}

\newcommand{\KeywordTok}[1]{\textcolor[rgb]{0.13,0.29,0.53}{\textbf{#1}}}
\newcommand{\NormalTok}[1]{#1}
\newcommand{\OperatorTok}[1]{\textcolor[rgb]{0.81,0.36,0.00}{\textbf{#1}}}

\newcommand{\StringTok}[1]{\textcolor[rgb]{0.31,0.60,0.02}{#1}}

\usepackage{longtable,booktabs}
\usepackage{graphicx,grffile}
\makeatletter
\def\maxwidth{\ifdim\Gin@nat@width>\linewidth\linewidth\else\Gin@nat@width\fi}
\def\maxheight{\ifdim\Gin@nat@height>\textheight\textheight\else\Gin@nat@height\fi}
\makeatother
\setkeys{Gin}{width=\maxwidth,height=\maxheight,keepaspectratio}
\IfFileExists{parskip.sty}{%
\usepackage{parskip}
}{
\setlength{\parindent}{0pt}
\setlength{\parskip}{6pt plus 2pt minus 1pt}
}
\providecommand{\tightlist}{%
  \setlength{\itemsep}{0pt}\setlength{\parskip}{0pt}}
\ifx\paragraph\undefined\else
\let\oldparagraph\paragraph
\renewcommand{\paragraph}[1]{\oldparagraph{#1}\mbox{}}
\fi
\ifx\subparagraph\undefined\else
\let\oldsubparagraph\subparagraph
\renewcommand{\subparagraph}[1]{\oldsubparagraph{#1}\mbox{}}
\fi

\let\rmarkdownfootnote\footnote%
\def\footnote{\protect\rmarkdownfootnote}

\usepackage{titling}

  \title{Eliciting the Endowment Effect under Assigned Ownership}
    \pretitle{\vspace{\droptitle}\centering\huge}
  \posttitle{\par}
    \author{Patrick Barranger, Rohit Nair, Rob Mulla, Shane Conner}
    \preauthor{\centering\large\emph}
  \postauthor{\par}
    \date{}
    \predate{}\postdate{}

\begin{document}
\maketitle

\hypertarget{abstract}{%
\subsection{Abstract}\label{abstract}}

In this study we present evidence that endowment effect can be elicited
merely by assigned ownership. Using Google Customer Survey, we
administered a survey were participants (n=495) were randomly split into
4 groups. Each group was assigned ownership of either legroom or their
ability to recline on an airline. Using this experiment setup we were
able to generate endowment effect, a 15-20x (at p\textless{}0.05)
increase between participant's willingness to pay (WTP) and their
willingness to accept (WTA).

\hypertarget{i.-introduction}{%
\subsection{I. Introduction}\label{i.-introduction}}

Many economic theories are developed under a rational agent model, where
consumers are expected to treat all information unbiasedly in their
decision making process. In the 1960s economists started to realize that
consumers did not always act in this anticipated rational manner.
Research lead to the development of behavioral economics, the study of
mental processes such as ``attention, language use, memory, perception,
problem solving, creativity, and thinking''{[}8{]}. One of the
hypothesis developed was the endowment effect, the hypothesis that
consumer ascribe more value to things merely because they own them. The
endowment effect captures the observation of the valuation paradigm,
where people will tend to pay more to retain something they own than to
obtain something they do not own --even when there is no cause for
attachment, or even if the item was only obtained minutes ago. Studies
of the effect typically focus on the difference in Willingness to Pay
(WTP) in comparison to the Willingness to Accept (WTA). In one famous
example participants were given a mug and then offered the chance to
sell or trade for an equally valued alternative, pens. The researchers
found that once participants had the mug their WTA was twice as high as
their WTP. In another similar study, researchers found that participants
selling price of NCAA final four tickets (WTA) was 14 times higher than
their WTP. With magnitudes ranging from two to 14 times in research, it
is clear that the endowment effect can cause people to assign
drastically different values due to simple possession. Loss aversion
(the disutility of giving up an object is greater that the utility
associated with acquiring it) and status quo bias (tendency of
individuals to remain in status quo than to leave it) {[}2{]}, were
initially shown to correlate with the endowment effect. Recent studies
have suggested that evolutionary, strategic and cognitive factors play a
part in eliciting endowment effect{[}4{]}.

Our study attempts to understand if an individual would respond
differently between a question where ownership is assigned and where it
is not. Our hypothesis is that assigning ownership will cause the
participant to price the object more highly. In this paper, we attempt
to answer this question by means of a questionnaire that sets up this
hypothetical ownership. This paper is organized as follows. Section II
describes our experiment design while in Section III we share data
analysis. Results and findings are discussed in section IV. Section V
concludes this paper.

\hypertarget{ii.-experimental-design}{%
\subsection{II. Experimental Design}\label{ii.-experimental-design}}

To test our hypothesis we set up a post-treatment measurement 2x2 test
design where we asked subjects to price seating features on a five hour
flight. Specifically, we asked them to either price legroom or the
recline feature. Half were asked their willingness to accept a payment
(WTA) to give up the feature and the other was asked the willingness to
pay (WTP) for the feature.

\hypertarget{data-gathering-and-subject-interaction}{%
\subsubsection{Data Gathering and Subject
Interaction}\label{data-gathering-and-subject-interaction}}

Our team decided to continue to use Google Consumer Surveys to gather
our data responses. Google surveys was chosen because of its relative
cost compared to other survey options, quick response time, and
additional information provided about the respondents like age, gender,
and location. Our team made a conscious decision to potentially trade
off quality of responses so that we could have more responses.

Our participants were mainly visitors of news websites who wanted to
read articles behind the website's paywall. Instead of paying, Google
allows customers to complete a surveys such as ours to read the article.
Each subject was asked only one of our four questions, and 133 responses
were gathered for each question. Participants could be reaching using
their computer or mobile device. We note that because most of our
respondents were attempting to get through paywalls they in a sense had
``self selected'' into our study. We also note that our sample
population may be similar in other ways due to this type of experiment,
specifically that they all are the type of people who desire to read
news articles.

\hypertarget{pilot-study}{%
\subsubsection{Pilot Study}\label{pilot-study}}

Before embarking on a full fledged study of endowment effect, we wanted
to ensure that different aspects of the study are working as expected.
To this effect we initiated a pilot study. Our object in running a pilot
study were:

\begin{itemize}
\tightlist
\item
  Ensuring Google Customer Survey (GCS) randomization works,
\item
  Survey takers understand our survey questions,
\item
  We are able to deliver treatment as we intended to,
\item
  Verify that responses are valid and meets expectation,
\item
  Distill presence of Endowment effect and
\item
  Get a baseline to calculate statistical power required for full
  fledged study.
\end{itemize}

The pilot study ran on Google Customer Survey from 25th June - 27 June,
2017. The survey consisted of two questions.

\begin{longtable}[]{@{}ll@{}}
\toprule
\begin{minipage}[b]{0.22\columnwidth}\raggedright
Question\strut
\end{minipage} & \begin{minipage}[b]{0.72\columnwidth}\raggedright
Description\strut
\end{minipage}\tabularnewline
\midrule
\endhead
\begin{minipage}[t]{0.22\columnwidth}\raggedright
1\strut
\end{minipage} & \begin{minipage}[t]{0.72\columnwidth}\raggedright
You are traveling on a 5 hour flight. What is the maximum amount you
would be willing to pay to recline your seat?\strut
\end{minipage}\tabularnewline
\begin{minipage}[t]{0.22\columnwidth}\raggedright
2\strut
\end{minipage} & \begin{minipage}[t]{0.72\columnwidth}\raggedright
You are traveling on a 5 hour flight. What is the minimum amount you
would accept to not recline your seat?\strut
\end{minipage}\tabularnewline
\bottomrule
\end{longtable}

Each survey taker is randomly assigned to one of the two questions and
they are provided with a textbox to post their response. The textbox
doesn't run any validation on the responses and accepts free form text.
This was done so as to gauge the kind of responses we get. We set the
survey to end when GCS receives at least 50 responses per question. Both
the questions assigns the respondent ownership of the chair. Question 1
is testing for their willingness to pay while question 2 tests their
willingness to accept. Once the survey ends, GCS provides functionality
to export survey responses with additional attributes pertaining to
respondents. The data consists of following columns:

\begin{longtable}[]{@{}ll@{}}
\toprule
\begin{minipage}[b]{0.22\columnwidth}\raggedright
Data Column\strut
\end{minipage} & \begin{minipage}[b]{0.72\columnwidth}\raggedright
Description\strut
\end{minipage}\tabularnewline
\midrule
\endhead
\begin{minipage}[t]{0.22\columnwidth}\raggedright
User ID\strut
\end{minipage} & \begin{minipage}[t]{0.72\columnwidth}\raggedright
A unique user ID for each survey respondent\strut
\end{minipage}\tabularnewline
\begin{minipage}[t]{0.22\columnwidth}\raggedright
Time (UTC)\strut
\end{minipage} & \begin{minipage}[t]{0.72\columnwidth}\raggedright
The time in which the survey was completed\strut
\end{minipage}\tabularnewline
\begin{minipage}[t]{0.22\columnwidth}\raggedright
Publisher Category\strut
\end{minipage} & \begin{minipage}[t]{0.72\columnwidth}\raggedright
The type of website the user was trying to view when filling out the
survey\strut
\end{minipage}\tabularnewline
\begin{minipage}[t]{0.22\columnwidth}\raggedright
Gender\strut
\end{minipage} & \begin{minipage}[t]{0.72\columnwidth}\raggedright
Gender of survey respondent or Unknown\strut
\end{minipage}\tabularnewline
\begin{minipage}[t]{0.22\columnwidth}\raggedright
Age\strut
\end{minipage} & \begin{minipage}[t]{0.72\columnwidth}\raggedright
The age bracket of the survey respondent or Unknown\strut
\end{minipage}\tabularnewline
\begin{minipage}[t]{0.22\columnwidth}\raggedright
Geography\strut
\end{minipage} & \begin{minipage}[t]{0.72\columnwidth}\raggedright
The Country, region and State of the respondent\strut
\end{minipage}\tabularnewline
\begin{minipage}[t]{0.22\columnwidth}\raggedright
Urban Density\strut
\end{minipage} & \begin{minipage}[t]{0.72\columnwidth}\raggedright
Population density of user's location (Urban, suburban, rural)\strut
\end{minipage}\tabularnewline
\begin{minipage}[t]{0.22\columnwidth}\raggedright
Income\strut
\end{minipage} & \begin{minipage}[t]{0.72\columnwidth}\raggedright
Income bracket user falls into\strut
\end{minipage}\tabularnewline
\begin{minipage}[t]{0.22\columnwidth}\raggedright
Parental Status\strut
\end{minipage} & \begin{minipage}[t]{0.72\columnwidth}\raggedright
Parental status of the user\strut
\end{minipage}\tabularnewline
\begin{minipage}[t]{0.22\columnwidth}\raggedright
Question \#\# Response\strut
\end{minipage} & \begin{minipage}[t]{0.72\columnwidth}\raggedright
Raw response of the respondent\strut
\end{minipage}\tabularnewline
\begin{minipage}[t]{0.22\columnwidth}\raggedright
Response Time \#1 (ms)\strut
\end{minipage} & \begin{minipage}[t]{0.72\columnwidth}\raggedright
Time taken to respond\strut
\end{minipage}\tabularnewline
\bottomrule
\end{longtable}

Once we exported the data, we loaded it in R and checked the validity of
the responses. Of the 101 responses, following are the non-numeric ones.
All but three (``Ly'', ``Cff'', ``5 hours'') of them can be easily
converted to numeric format. We will mark these three as NA so as not to
affect our calculations.

Odd responses

\begin{longtable}[]{@{}ll@{}}
\toprule
Odd & Responses\tabularnewline
\midrule
\endhead
Ly & \$2\tabularnewline
\$50 & \$10\tabularnewline
\$0 & 0\ldots{}I think it's rude to recline on any
flight\ldots{}ever\tabularnewline
\$100 & 50 dollars\tabularnewline
Cff & \$100\tabularnewline
5 hours & \$100\tabularnewline
\$150 & 300.00 us\tabularnewline
Zero & 1,000000000.00\tabularnewline
\bottomrule
\end{longtable}

Following this we plot parsed responses on a box plot to see wether the
distribution is as expected. We see that the median response for
question 1 is around \$10 while that for question 2 is around \$100. The
difference between responses to these two questions is the endowment
effect which we were able to elicit in our pilot study which is
promising.

\includegraphics{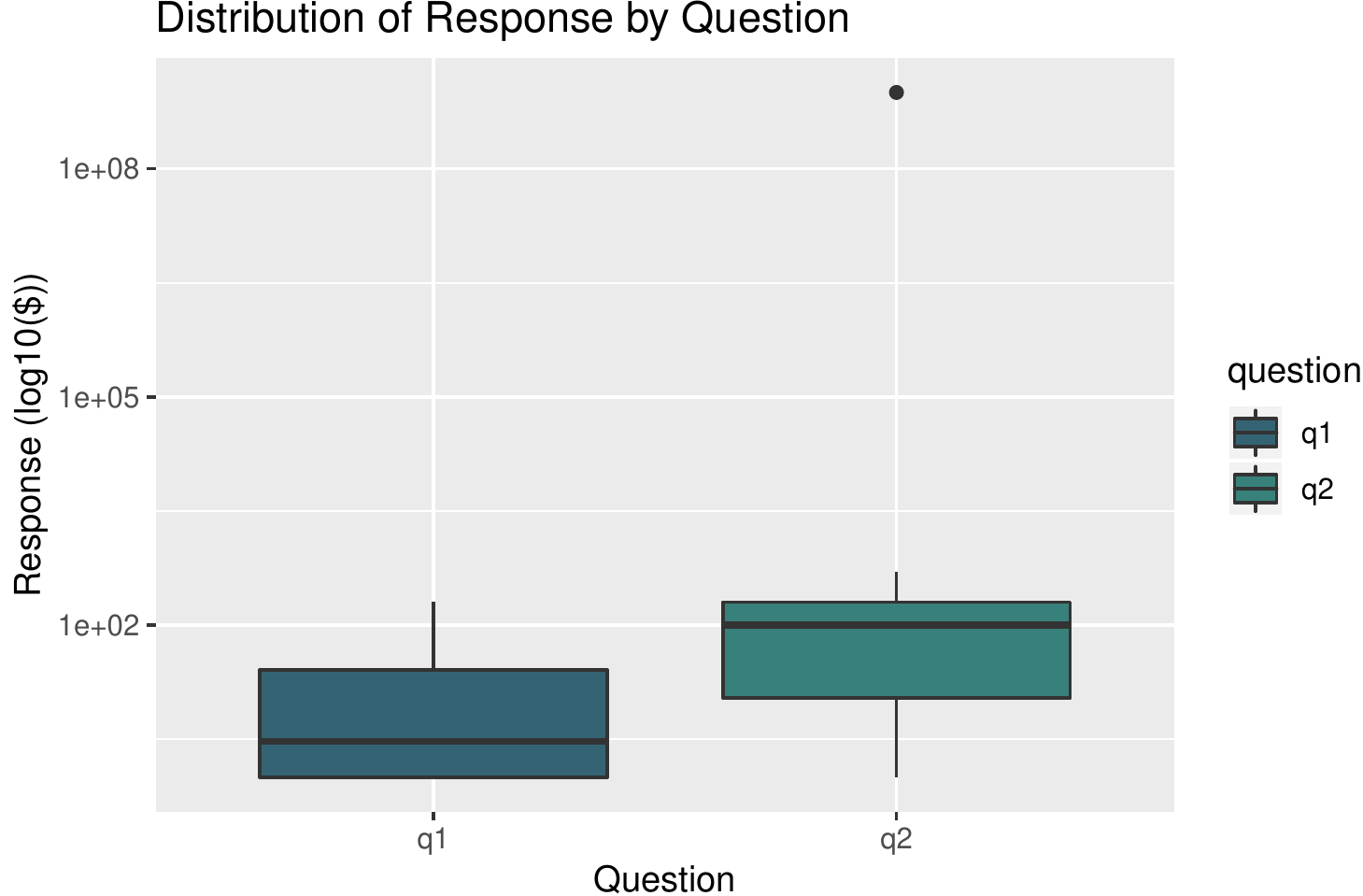}

Next we check wether the responses are distributed uniformly among
different demographics. Below we show distribution of responses for each
question by Gender/Age/Income.

\includegraphics{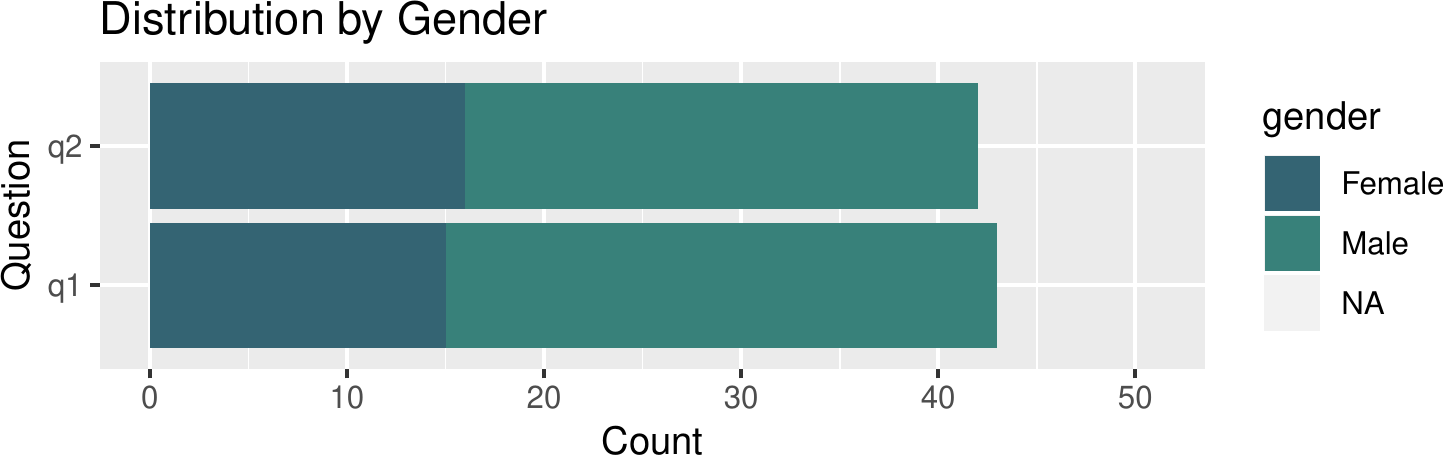}

\includegraphics{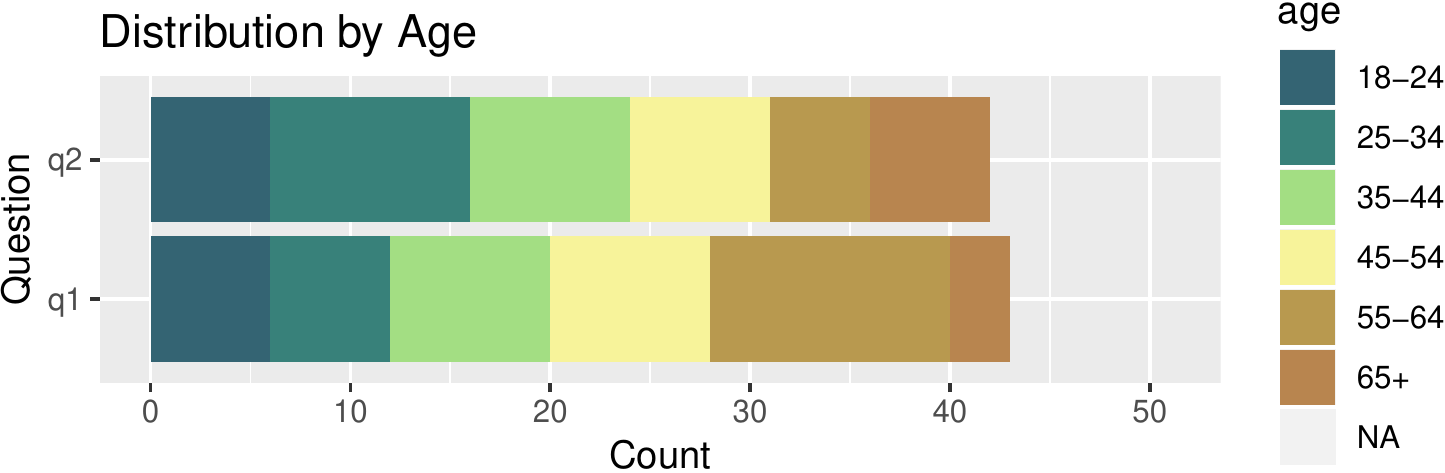}

\includegraphics{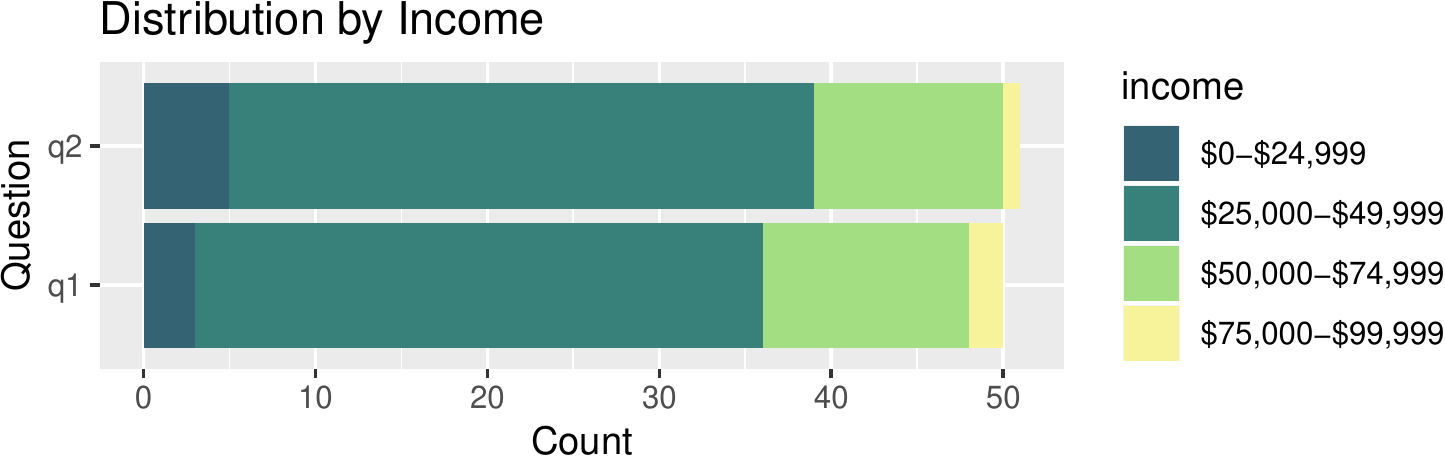}

Next, we ran a test for covariate balance. We create a subset of
responses by omitting NAs for the covariates. We create two regression
models, one with all the covariates and another with no covariates.
Using ANOVA we test whether any of these

\begin{verbatim}
## Analysis of Deviance Table
## 
## Model 1: treatment ~ gender + region + age
## Model 2: treatment ~ 1
##   Resid. Df Resid. Dev Df Deviance Pr(>Chi)
## 1        75     18.435                     
## 2        84     21.247 -9  -2.8117   0.2468
\end{verbatim}

The result is a non statistically significant p-value of 2468 which
suggests the randomization design was not violated. The data from the
pilot study shows that we were able to get right responses for our
questions and that GCS did a good job of randomizing respondents.

\hypertarget{actual-study}{%
\subsubsection{Actual Study}\label{actual-study}}

In our actual study we expanded our questions to four, gathering 133
responses per question. These questions were:

\begin{longtable}[]{@{}ll@{}}
\toprule
\begin{minipage}[b]{0.09\columnwidth}\raggedright
Question Number\strut
\end{minipage} & \begin{minipage}[b]{0.85\columnwidth}\raggedright
Question Text\strut
\end{minipage}\tabularnewline
\midrule
\endhead
\begin{minipage}[t]{0.09\columnwidth}\raggedright
q1\strut
\end{minipage} & \begin{minipage}[t]{0.85\columnwidth}\raggedright
On a 5 hour flight what is the maximum amount you would be willing to
pay to recline the seat?\strut
\end{minipage}\tabularnewline
\begin{minipage}[t]{0.09\columnwidth}\raggedright
q2\strut
\end{minipage} & \begin{minipage}[t]{0.85\columnwidth}\raggedright
On a 5 hour flight what is the minimum amount you would accept to not
recline your seat?\strut
\end{minipage}\tabularnewline
\begin{minipage}[t]{0.09\columnwidth}\raggedright
q3\strut
\end{minipage} & \begin{minipage}[t]{0.85\columnwidth}\raggedright
On a 5 hour flight what is the maximum amount you would be willing to
pay to stop the passenger in front of you from reclining into the space
in front of you?\strut
\end{minipage}\tabularnewline
\begin{minipage}[t]{0.09\columnwidth}\raggedright
q4\strut
\end{minipage} & \begin{minipage}[t]{0.85\columnwidth}\raggedright
On a 5 hour flight what is the minimum amount you would accept to allow
the passenger in front of you to recline into your space?\strut
\end{minipage}\tabularnewline
\bottomrule
\end{longtable}

\newpage

\hypertarget{iii.-data-analysis}{%
\subsection{III. Data Analysis}\label{iii.-data-analysis}}

Our raw data is provided to us by google surveys in the form of a xls
file. White our main concern is the answers provided by the survey
respondents, we are also provided additional information about our
sample population. We can run some statistical tests to determine if
these are the same for each group.

\begin{longtable}[]{@{}ll@{}}
\toprule
\begin{minipage}[b]{0.22\columnwidth}\raggedright
Data Column\strut
\end{minipage} & \begin{minipage}[b]{0.72\columnwidth}\raggedright
Description\strut
\end{minipage}\tabularnewline
\midrule
\endhead
\begin{minipage}[t]{0.22\columnwidth}\raggedright
User ID\strut
\end{minipage} & \begin{minipage}[t]{0.72\columnwidth}\raggedright
A unique user ID for each survey respondent\strut
\end{minipage}\tabularnewline
\begin{minipage}[t]{0.22\columnwidth}\raggedright
Time (UTC)\strut
\end{minipage} & \begin{minipage}[t]{0.72\columnwidth}\raggedright
The time in which the survey was completed\strut
\end{minipage}\tabularnewline
\begin{minipage}[t]{0.22\columnwidth}\raggedright
Publisher Category\strut
\end{minipage} & \begin{minipage}[t]{0.72\columnwidth}\raggedright
The type of website the user was trying to view when filling out the
survey\strut
\end{minipage}\tabularnewline
\begin{minipage}[t]{0.22\columnwidth}\raggedright
Gender\strut
\end{minipage} & \begin{minipage}[t]{0.72\columnwidth}\raggedright
Gender of survey respondent or Unknown\strut
\end{minipage}\tabularnewline
\begin{minipage}[t]{0.22\columnwidth}\raggedright
Age\strut
\end{minipage} & \begin{minipage}[t]{0.72\columnwidth}\raggedright
The age bracket of the survey respondent or Unknown\strut
\end{minipage}\tabularnewline
\begin{minipage}[t]{0.22\columnwidth}\raggedright
Geography\strut
\end{minipage} & \begin{minipage}[t]{0.72\columnwidth}\raggedright
The Country, region, State and City of the respondent\strut
\end{minipage}\tabularnewline
\begin{minipage}[t]{0.22\columnwidth}\raggedright
Question Raw Response\strut
\end{minipage} & \begin{minipage}[t]{0.72\columnwidth}\raggedright
Raw response of the respondent\strut
\end{minipage}\tabularnewline
\begin{minipage}[t]{0.22\columnwidth}\raggedright
Response Time \#1 (ms)\strut
\end{minipage} & \begin{minipage}[t]{0.72\columnwidth}\raggedright
Time taken to respond\strut
\end{minipage}\tabularnewline
\bottomrule
\end{longtable}

\hypertarget{response-filtering-and-noncompliance}{%
\subsubsection{Response filtering and
Noncompliance}\label{response-filtering-and-noncompliance}}

Our survey allowed the participant to type any text in the response box.
The text box, before the respondents start entering text stated ``Enter
your answer in US \$''. Because of this the responses were not all
easily converted into numerical values. Some manual interpretation was
required. Many of the responses could be converted, for instance `2
dollars' could be interpreted as `2'. Other responses such as those in
units of measurement (3 inches), units of time (3 hours) or nonsense
responses (wa) were all converted to NA values and ignored for our
analysis. These participants are considered `noncompliers' in our study.

Below are the responses that were manually converted to numeric values.

\begin{Shaded}
\begin{Highlighting}[]
\CommentTok{# Cases which can be turned into numerical values}
\NormalTok{d[d}\OperatorTok{$}\NormalTok{response }\OperatorTok{==}\StringTok{ "none it should be free"}\NormalTok{]}\OperatorTok{$}\NormalTok{response <-}\StringTok{ }\DecValTok{0}
\NormalTok{d[d}\OperatorTok{$}\NormalTok{response }\OperatorTok{==}\StringTok{ "1 000 000 000.69"}\NormalTok{]}\OperatorTok{$}\NormalTok{response <-}\StringTok{ }\FloatTok{1000000000.69}
\NormalTok{d[d}\OperatorTok{$}\NormalTok{response }\OperatorTok{==}\StringTok{ "two"}\NormalTok{]}\OperatorTok{$}\NormalTok{response <-}\StringTok{ }\DecValTok{2}
\NormalTok{d[d}\OperatorTok{$}\NormalTok{response }\OperatorTok{==}\StringTok{ "zero"}\NormalTok{]}\OperatorTok{$}\NormalTok{response <-}\StringTok{ }\DecValTok{0}
\NormalTok{d[d}\OperatorTok{$}\NormalTok{response }\OperatorTok{==}\StringTok{ "nothing"}\NormalTok{]}\OperatorTok{$}\NormalTok{response <-}\StringTok{ }\DecValTok{0}
\NormalTok{d[d}\OperatorTok{$}\NormalTok{response }\OperatorTok{==}\StringTok{ "none"}\NormalTok{]}\OperatorTok{$}\NormalTok{response <-}\StringTok{ }\DecValTok{0}
\NormalTok{d[d}\OperatorTok{$}\NormalTok{response }\OperatorTok{==}\StringTok{ "195 500 812.50"}\NormalTok{]}\OperatorTok{$}\NormalTok{response <-}\StringTok{ }\FloatTok{195500812.50}
\NormalTok{d[d}\OperatorTok{$}\NormalTok{response }\OperatorTok{==}\StringTok{ "100 000 000.836215"}\NormalTok{]}\OperatorTok{$}\NormalTok{response <-}\StringTok{ }\FloatTok{100000000.836215}
\NormalTok{d[d}\OperatorTok{$}\NormalTok{response }\OperatorTok{==}\StringTok{ "non"}\NormalTok{]}\OperatorTok{$}\NormalTok{response <-}\StringTok{ }\DecValTok{0}
\NormalTok{d[d}\OperatorTok{$}\NormalTok{response }\OperatorTok{==}\StringTok{ "10 000"}\NormalTok{]}\OperatorTok{$}\NormalTok{response <-}\StringTok{ }\DecValTok{10000}
\NormalTok{d[d}\OperatorTok{$}\NormalTok{response }\OperatorTok{==}\StringTok{ "0.00 as long as the seat reclines it is their right to recline it"}\NormalTok{]}\OperatorTok{$}\NormalTok{response <-}\StringTok{ }\DecValTok{0}
\NormalTok{d[d}\OperatorTok{$}\NormalTok{response }\OperatorTok{==}\StringTok{ "25 dollars"}\NormalTok{]}\OperatorTok{$}\NormalTok{response <-}\StringTok{ }\DecValTok{25}
\NormalTok{d[d}\OperatorTok{$}\NormalTok{response }\OperatorTok{==}\StringTok{ "zedo"}\NormalTok{]}\OperatorTok{$}\NormalTok{response <-}\StringTok{ }\DecValTok{0}
\NormalTok{d[d}\OperatorTok{$}\NormalTok{response }\OperatorTok{==}\StringTok{ "1 000 000.00"}\NormalTok{]}\OperatorTok{$}\NormalTok{response <-}\StringTok{ }\FloatTok{1000000.00}
\NormalTok{d}\OperatorTok{$}\NormalTok{response <-}\StringTok{ }\KeywordTok{as.numeric}\NormalTok{(d}\OperatorTok{$}\NormalTok{response)}
\end{Highlighting}
\end{Shaded}

The following were responses that were converted to \texttt{NA}.

\begin{verbatim}
##  [1] "i don't know"           "i don't fly commercial"
##  [3] "2 hours"                "reject offer"          
##  [5] "250 free flight"        "2 hours"               
##  [7] "2 hours"                "2 min"                 
##  [9] "do not follow question" "na"                    
## [11] "yes"                    "wa"                    
## [13] "2.5 hour"               "not at all"            
## [15] "yes"                    "3 hours"               
## [17] "5 hours"                "15 minutes"            
## [19] "5 hours"                "1 hour"                
## [21] "not sure"               "3 inches"
\end{verbatim}

We can plot a count by question to inspect noncompliance by question. We
note that visually it appears that the distribution of NA responses is
not even between questions. The `Willingness to Pay' questions (Question
\#1 and Question \#3) have fewer NA responses compared to our
`Willingness to Accept' questions (Question \#2 and Question \#4). This
may be attributed to the fact that the willingness to accept questions
were fundamentally harder to understand, which may have impacted our
randomization and must be accounted for in our final analysis.

\includegraphics{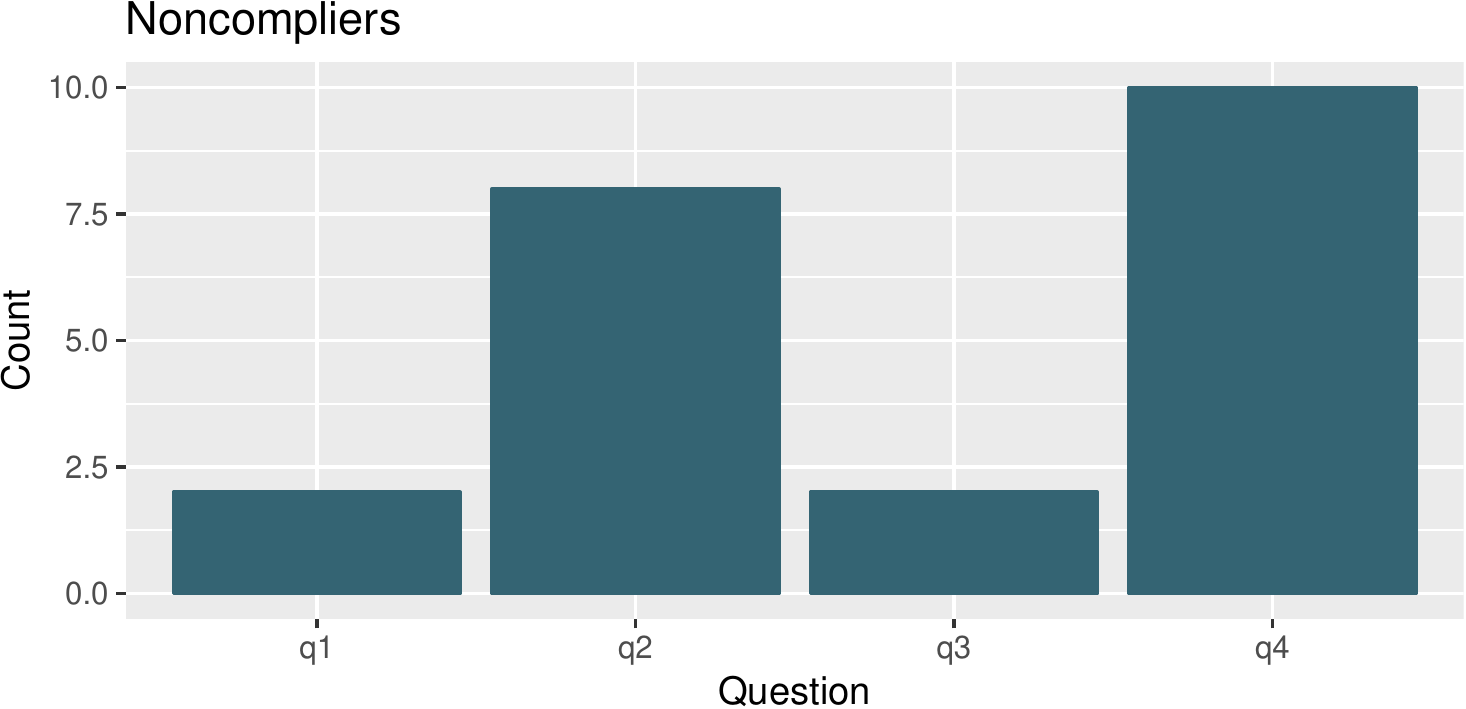}

Additionally we conduct a \texttt{t.test} and \texttt{cohen.d} test to
compare the compliance rate of our treatment and control groups. The
results of the t-test show a statistically significant difference
between the compliance rate while the Cohen's D shows a small effect
size between the two groups. In this study we are mainly concerned with
the complier average causal effect (CACE). Simply looking at the
intention to treat (ITT) would give these nonsense responses some
weight, while the CACE give us the effect for those who actually
understood the question and answered the question in compliance with our
survey request of ``Enter your answer in US \$''.

\begin{verbatim}
## 
##  Welch Two Sample t-test
## 
## data:  d$compliance[d$treatment == 1] and d$compliance[d$treatment == 0]
## t = -3.0802, df = 382.93, p-value = 0.002218
## alternative hypothesis: true difference in means is not equal to 0
## 95 percent confidence interval:
##  -0.08608726 -0.01900442
## sample estimates:
## mean of x mean of y 
## 0.9325843 0.9851301
\end{verbatim}

\begin{verbatim}
## 
## Cohen's d
## 
## d estimate: -0.2667107 (small)
## 95 percent confidence interval:
##         inf         sup 
## -0.43716436 -0.09625695
\end{verbatim}

\hypertarget{check-for-randomization}{%
\subsubsection{Check for Randomization}\label{check-for-randomization}}

To check for randomization we will compare the responses between
questions. If randomization was done correctly there should be no
statistically significant difference between the response rates for each
question. First we will look at the results visually.

The distribution of gender per question shows a slightly less number of
known female respondents for all questions except question \#1.

\includegraphics{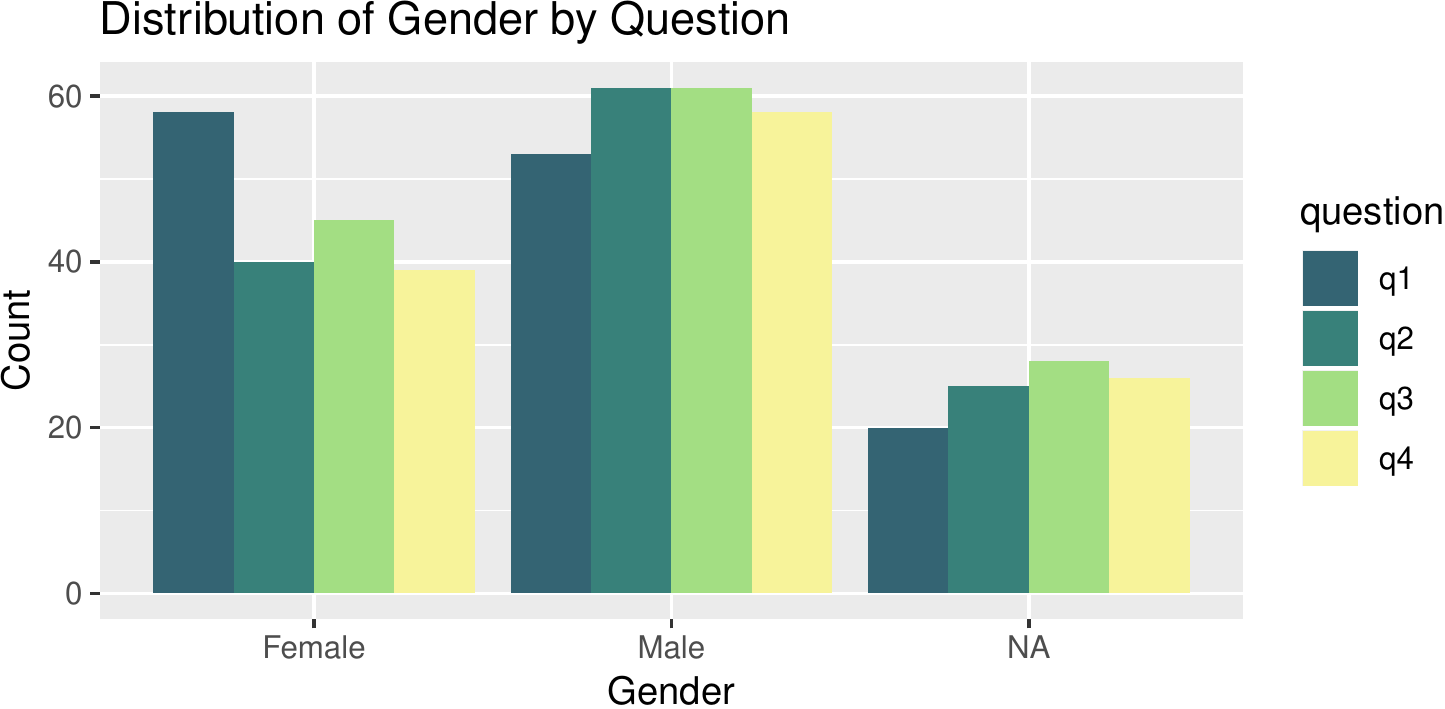}

We can compare how long it took respondents to complete the question by
using a boxplot. Note the y-axis is on a logarithmic scale. Question 2
appears to be the only question with a slightly longer average response
time.

\includegraphics{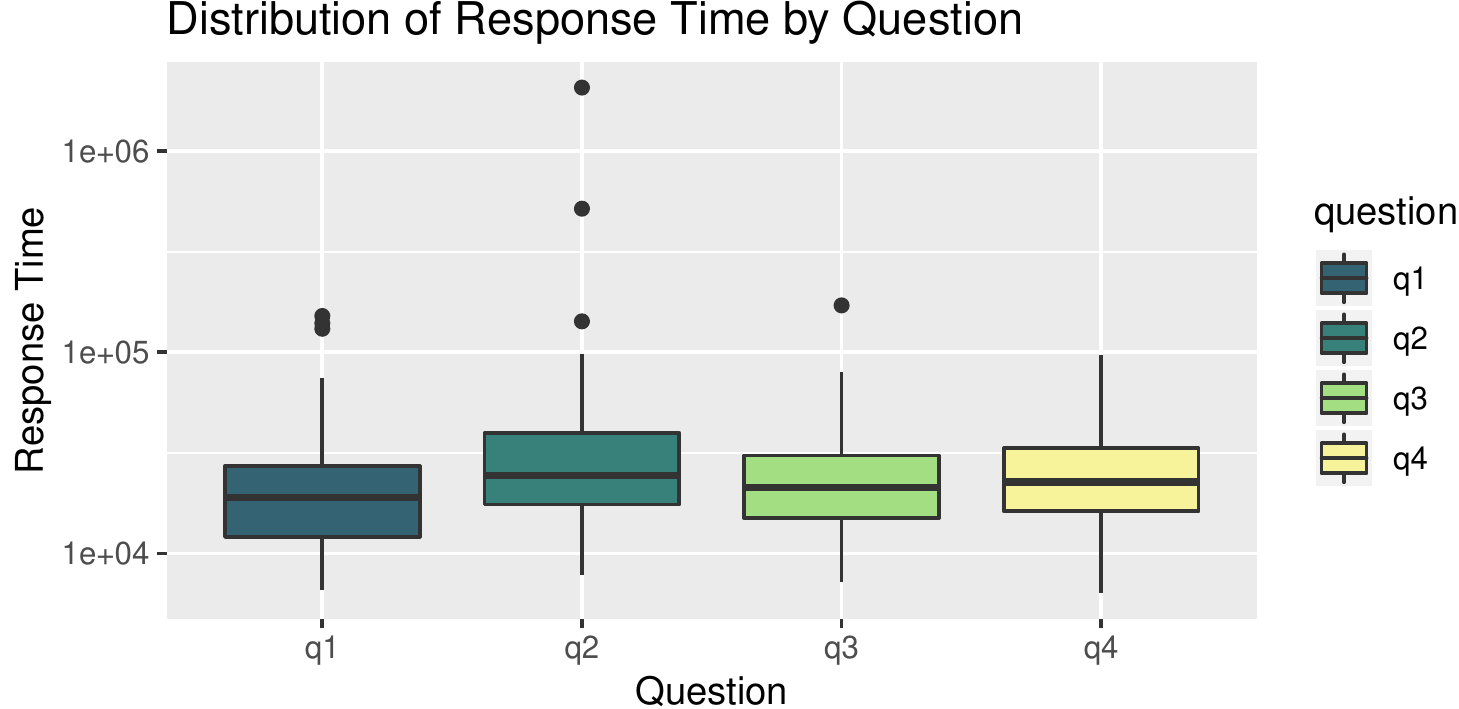}

Exploring the age distribution of respondents we see no obvious bias.
The number of respondents in the 18-24 age bracket appears to have the
largest difference between questions. With the small number of
respondents we cannot say this is statistically significant.

\includegraphics{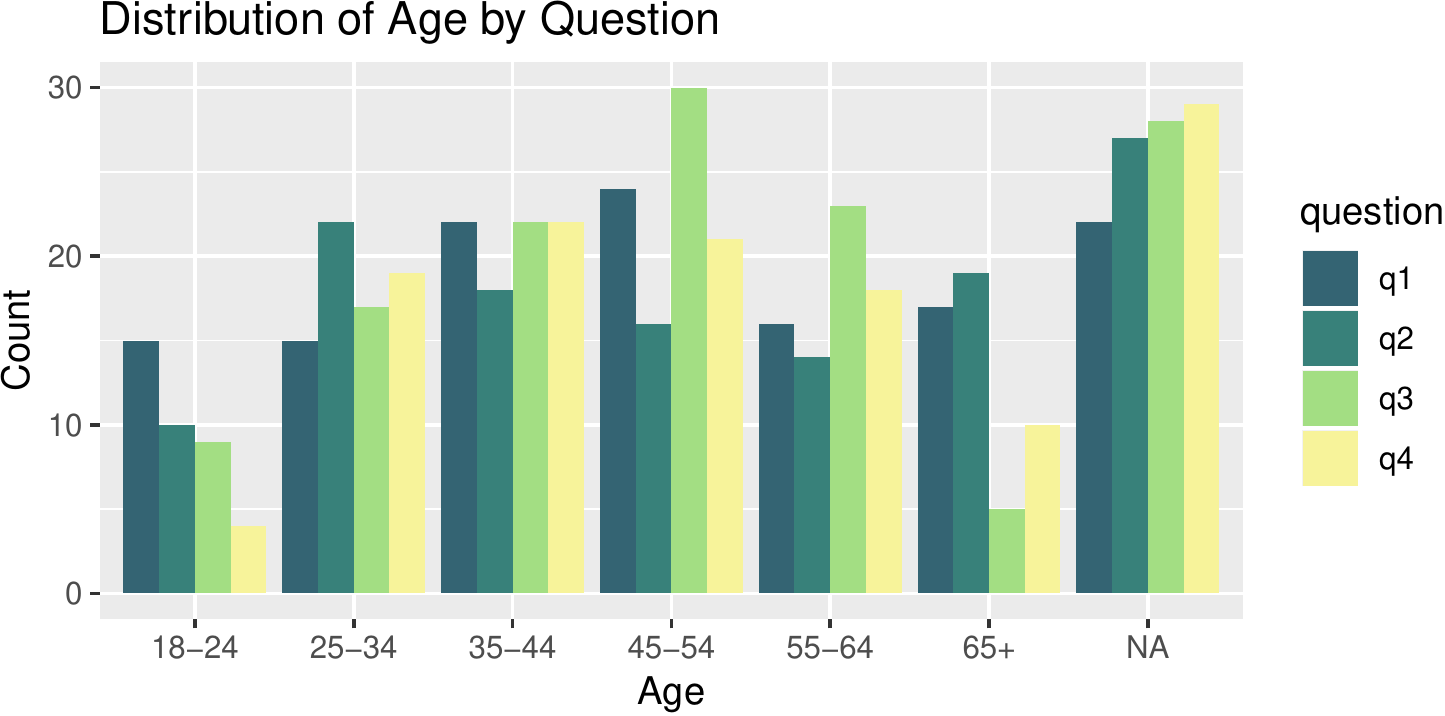}

\hypertarget{anova-test}{%
\subsubsection{ANOVA test}\label{anova-test}}

Next, we create a subset from our dataset and remove any observations
that do not include the categorical data we are leveraging to check
covariate balance.

We isolate the first question set's observations and check if any
categorical factors had an effect on whether treatment was assigned.

\begin{verbatim}
## Analysis of Deviance Table
## 
## Model 1: treatment ~ gender + region + age
## Model 2: treatment ~ 1
##   Resid. Df Resid. Dev  Df Deviance Pr(>Chi)
## 1       199     278.43                      
## 2       209     290.44 -10  -12.006   0.2847
\end{verbatim}

The result is a non statistically significant p-value of 0.3341 which
suggests the randomization design was not violated.

We repeat this on the second test:

\begin{verbatim}
## Analysis of Deviance Table
## 
## Model 1: treatment ~ gender + region + age
## Model 2: treatment ~ 1
##   Resid. Df Resid. Dev  Df Deviance Pr(>Chi)
## 1       190     268.68                      
## 2       200     278.04 -10  -9.3636    0.498
\end{verbatim}

Which also results in a non statistically significant p-value of 0.4674.

\hypertarget{transformations}{%
\subsubsection{Transformations}\label{transformations}}

Due to the continuous user input there are a handful of extreme
outliers, particularly in the WTA pool (e.g., willing to accept
\$10,000+).

\includegraphics{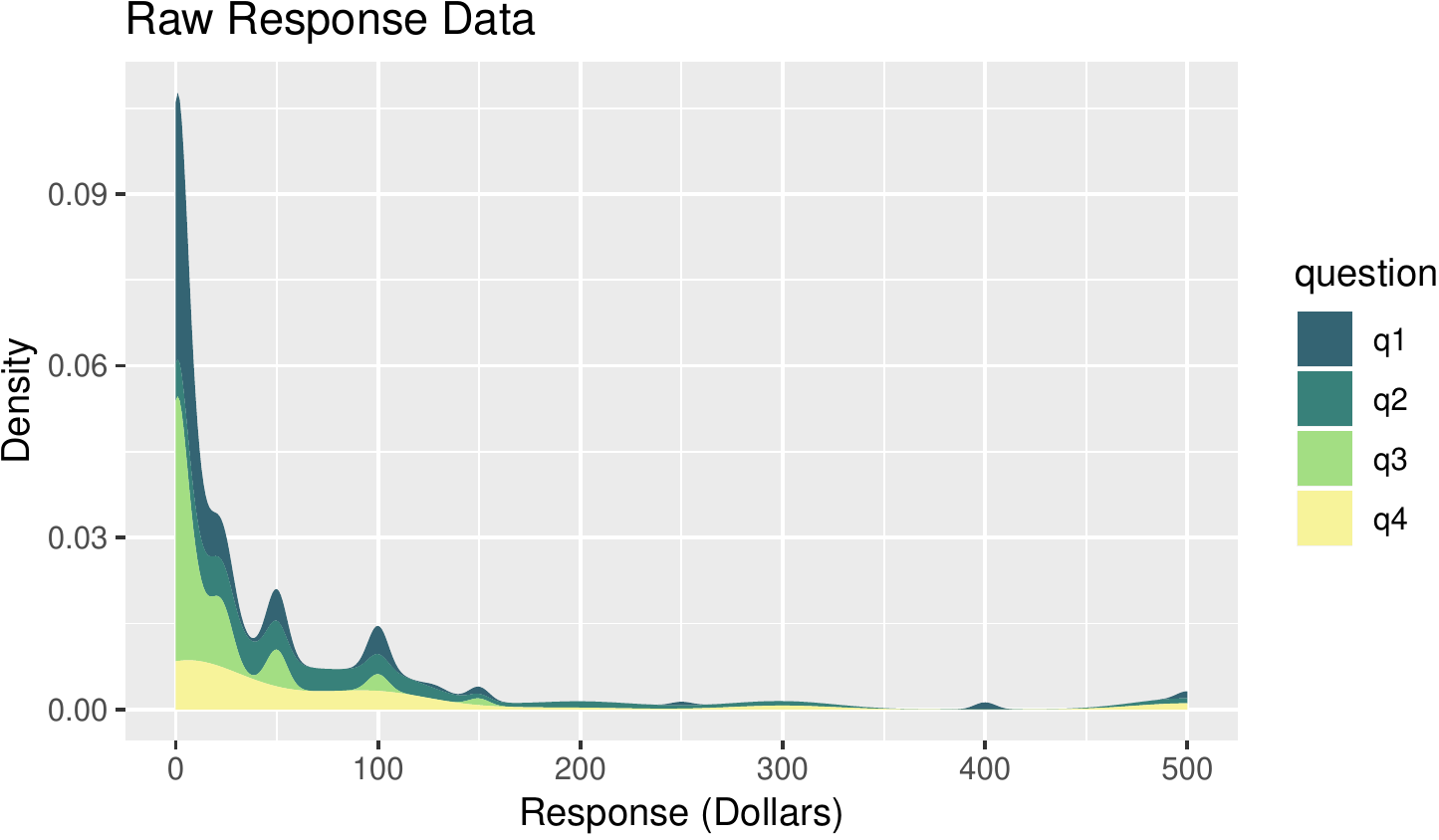}

The responses of the raw data resulted in an average user WTP as
\$9126012 (\$5 median) and WTA \$794316 (\$75 median) for the first
question set (recline). WTP as \$15.27 (\$1 median) and WTA \$8259.81
(\$20 median) for the second question set (legroom).

At first, we considered removing these observations altogether since
they are disruptive to the response mean and likely an unconsidered
response. And perhaps they are. However, after discussing amongst
ourselves we decided we ultimately will treat these responses as valid
data despite our reservations. For the purposes of our analysis, we are
concluding these users are unwilling to negotiate an accept value which
is a legitimate response. Although this decision compromises the
usefulness of our data as is. In light of these considerations, our data
required transformation.

To get a better sense of the response data relative to itself we
explored the data using nonparametric statistics.

Resultantly, a bimodal distribution was unveiled. Collectively, the
largest density of responses were subjects who answered \$0 which
resulted in one of the two humps. The second is due to the subjects in
the treatment (WTA) consistently responding in a higher max value than
their control (WTP) counterparts. The first question set identified the
average WTP subject in the 0.381 percentile vs the 0.632 percentile for
WTA subjects. 0.427 percentile for WTP and 0.595 percentile for WTA in
the second question set

In addition to better understanding the distribution, both question sets
were highly statistically significant using the nonparametric Wilcoxon
Rank-sum test model. Despite the promising results, our analysis still
left us without an answer for the average a subject is willing to pay in
comparison to what they're willing to accept.

To better answer that question, we used a log transformation on the
data.

\begin{Shaded}
\begin{Highlighting}[]
\CommentTok{# Stacked plot}
\KeywordTok{grid.arrange}\NormalTok{(log_recline, log_legroom, }\DataTypeTok{ncol=}\DecValTok{1}\NormalTok{, }\DataTypeTok{nrow=}\DecValTok{2}\NormalTok{)}
\end{Highlighting}
\end{Shaded}

\includegraphics{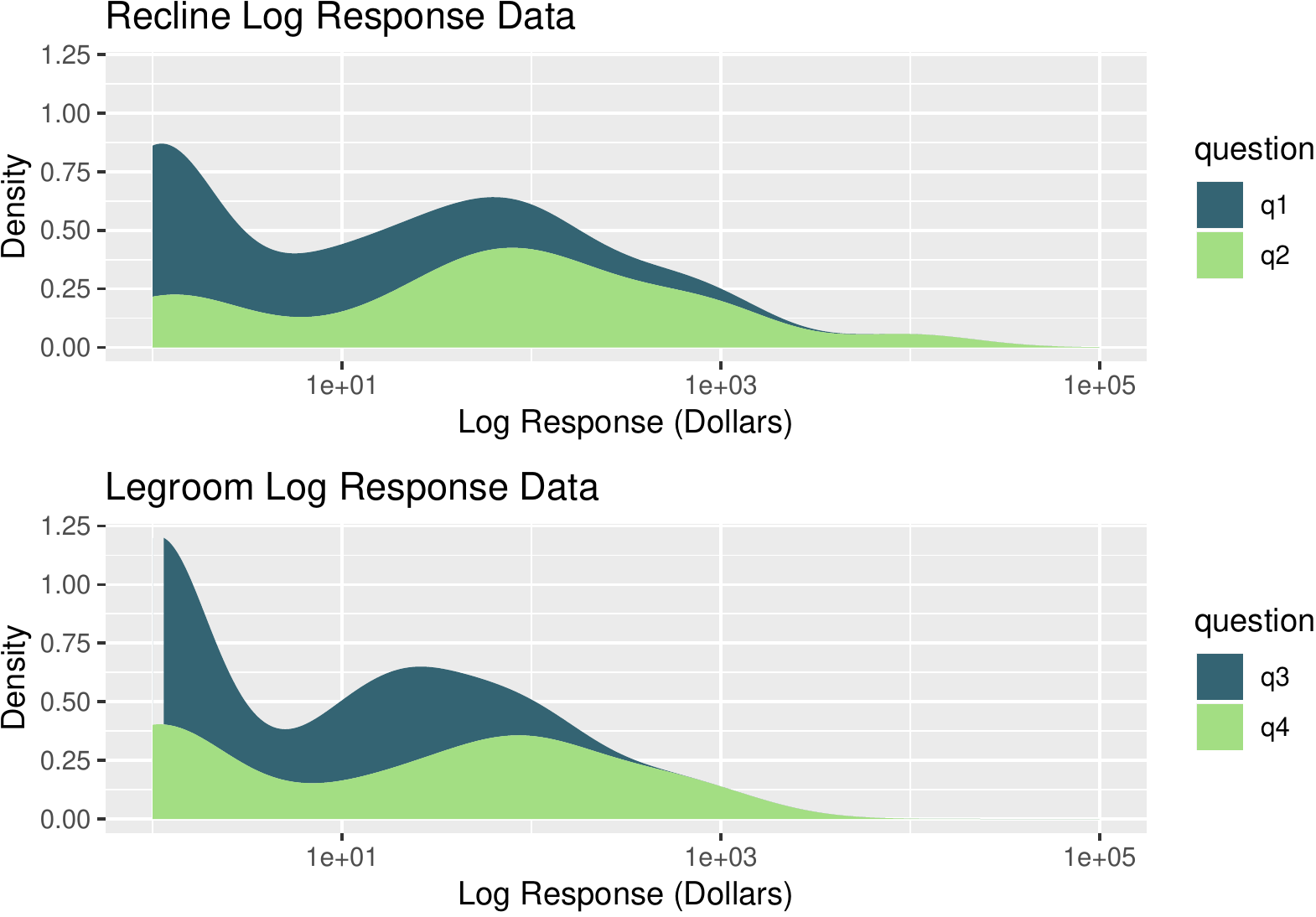}

\includegraphics{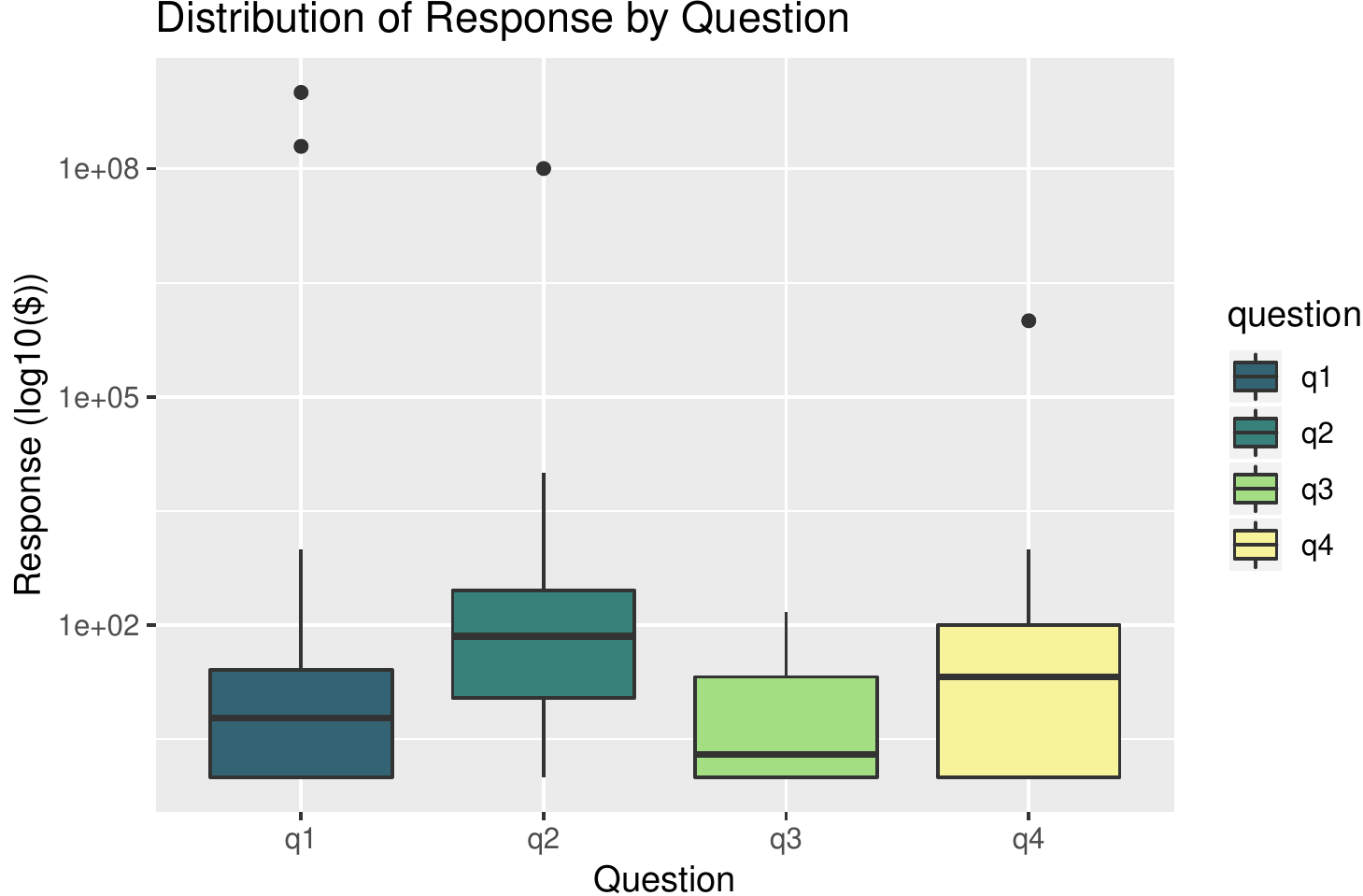}

\hypertarget{iv.-results-discussion}{%
\subsection{IV. Results \& Discussion}\label{iv.-results-discussion}}

Given the non-parametric distribution of our data we first analyzed our
data using a Wilcoxon rank sum test.

\begin{verbatim}
## 
##  Wilcoxon rank sum test with continuity correction
## 
## data:  d$response[(d$recline == 1) & (d$treatment == 1)] and d$response[(d$recline == 1) & (d$treatment == 0)]
## W = 12232, p-value = 1.247e-11
## alternative hypothesis: true location shift is not equal to 0
\end{verbatim}

\begin{verbatim}
## 
##  Wilcoxon rank sum test with continuity correction
## 
## data:  d$response[(d$recline == 0) & (d$treatment == 1)] and d$response[(d$recline == 0) & (d$treatment == 0)]
## W = 10858, p-value = 5.68e-06
## alternative hypothesis: true location shift is not equal to 0
\end{verbatim}

Our results showed that for both our recline and legroom question sets
we reject our null hypothesis that there is no shift in rank. These
results gave us evidence to reject the null hypothesis that assigned
ownership does not have an impact on the value of an object, but failed
to show the magnitude of the effect.

We wanted to quantify the amount of difference observed between subjects
beyond simply stating their rank was different. Given the skewed
distribution of our observations and the large difference between the
mean and median in each respondent group we decided to analyze our
results using randomization inference on the median response.

\includegraphics{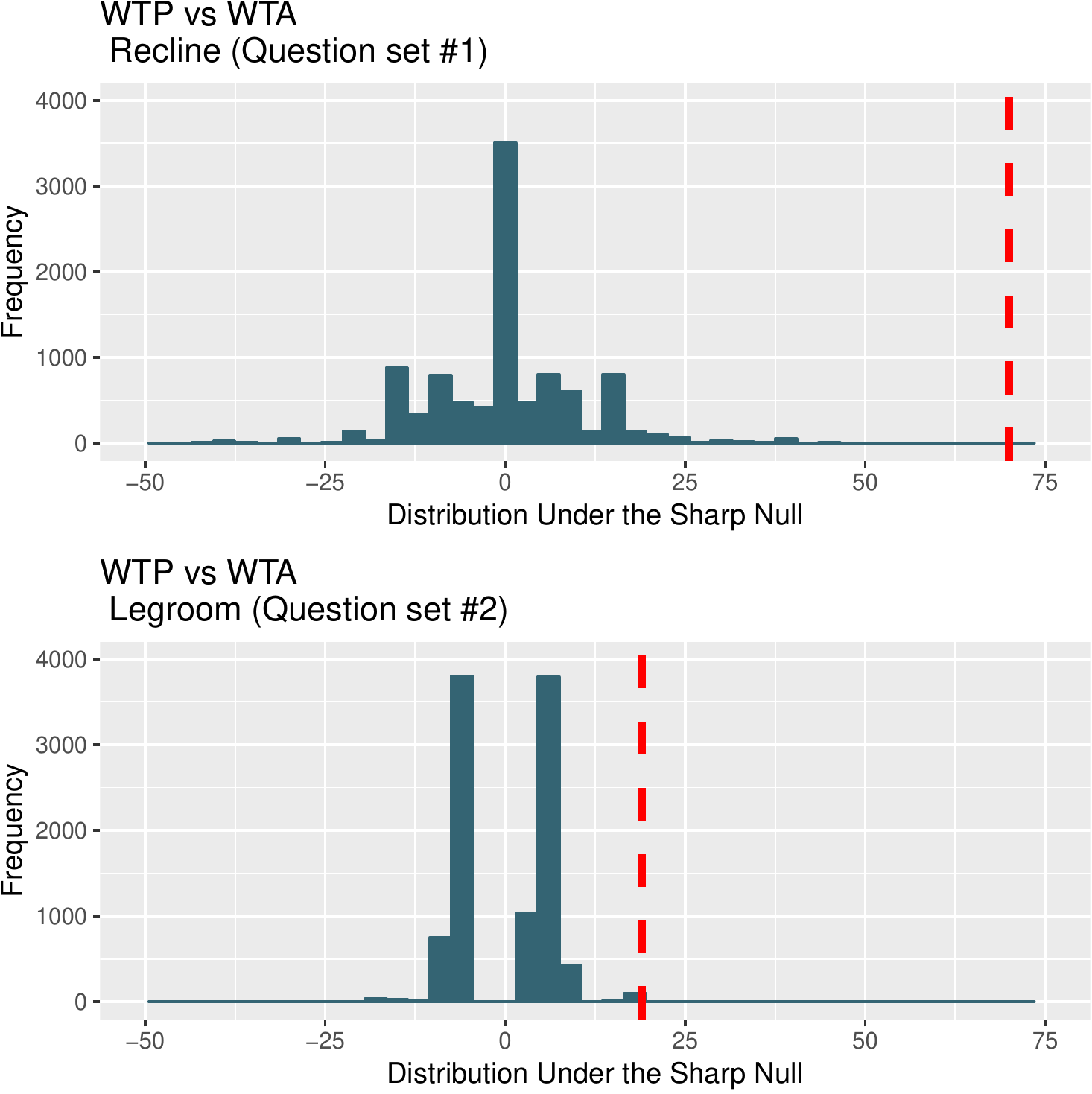}

Subjects who answered our WTPxRecline question (q1) had a median
response of (\$5) and our WTAxRecline (q2) subjects had a median
response of (\$75). The difference in median of \$70 (15:1 WTA:WTP) is
statistically significant with 0 of 10000 draws showing a larger
difference (p value of 0).

Similar to our recline question set, subjects who answered our
WTPxLegroom question (q3) had a median response of \$1 and our
WTAxLegroom (q4) subjects had a median response of (\$20). The
difference in median of \$19 (20:1 WTA:WTP) is statistically significant
with 40 of 10000 draws showing a larger difference (p value of 0.004).

Given the statistically significant results of both the recline and
legroom question sets we can reject our null hypothesis that assigned
ownership does not have an impact on the value of an object. We observed
that the ratio of WTA:WTP ranged from 15:1 for reclining and 20:1 for
legroom.

Our 2x2 design allows us to set up hypothetical pairs of real world
`negotiations' between a recliner and the person behind them. We used
the same Randomization Inference methodology for the medians of q1 vs q4
(WTPxRecline vs WTAxLegroom) and q2 vs q3 (WTAxRecline vs WTPxLegroom).
As the endowment effect would predict, we observe that the WTA
individuals have a statistically significant higher median minimum
amount required than the median WTP. WTAxlegroom individuals wanted a
median minimum payment of 20 while the WTPxrecline individuals wanted to
spend a median maximum payment of 5. This difference is statistically
significant with a p-value of 0.0234 at an observed 4:1 WTA:WTP.
WTAxrecline individuals wanted a median minimum payment of 75 while the
WTPxlegroom individuals wanted to spend a median maximum payment of 1.
This difference is statistically significant with a p-value of 0 at an
observed 75:1 WTA:WTP. This has interesting real-world implications in
that the object, legroom/reclining in this study, has different
negotiated values depending on who is given the ownership - even though
rational-agent economic theory suggests individuals should value the
object identically!

\hypertarget{v.-conclusion}{%
\subsection{V. Conclusion}\label{v.-conclusion}}

Our results show that we can reject our null hypothesis that assigned
ownership does not have an impact on the value of an object. While this
is exciting, we need to be cautious about how far to extend the
implications of these results. There does not exist a standing framework
with airlines to allow passengers to negotiate seat privileges. We would
anticipate real world tests to have slightly different outcomes than our
hypothetical negotiation tests due to additional behavioral economic
interactions, such as price setting (from overheard negotiations). That
said, the intent for this study was to evaluate how assigned ownership
in a question causes an individual to price an object differently, not
how to help airline best set seat pricing.

The statistically significant results we observed have implications
outside of the airline industry. For example, voters are often asked
through ballot measures to vote on approval of public work projects or
government services. For many voters this decision includes determining
in their mind what the price of the good or service is and if the
proposal is an acceptable deal. Our research would indicate that the way
the question on the ballot is posed could have a material impact on the
outcome. Consistent with our results, we would anticipate voters to have
a higher assumed value to objects or services they perceive to ``own'',
such as a park or river. Just like the reclining seat in our study,
these objects or services on ballots rarely have a known market price.
Moreover just like our subjects, voters are simply given a question and
rarely involved in the negotiation to get to the ballot-stated value.
While our results lack a real world mechanism for seat negotiations they
have strong implications for similar real-world situations where
individuals are asked to determine the value of a good or service.

Our work gives strong evidence that assigned ownership changes how an
individual perceives the worth of an object. While we can't extend these
results too far, future research into how an individual comes to believe
ownership would be interesting. Additionally, given our positive result
it would be interesting to see how assigned ownership interacts with
political beliefs and if this effect extends into how individuals vote.

\hypertarget{references}{%
\subsection{References}\label{references}}

\begin{enumerate}
\def\labelenumi{\arabic{enumi}.}
\tightlist
\item
  Kahneman, Daniel, Jack L. Knetsch, and Richard H. Thaler.
  ``Experimental tests of the endowment effect and the Coase theorem.''
  Journal of political Economy 98.6 (1990): 1325-1348.
\item
  Kahneman, Daniel, Jack L. Knetsch, and Richard H. Thaler. ``Anomalies:
  The endowment effect, loss aversion, and status quo bias.'' The
  journal of economic perspectives 5.1 (1991): 193-206.
\item
  Ericson, Keith M. Marzilli, and Andreas Fuster. ``The endowment
  effect.'' (2014).
\item
  Morewedge, Carey K., and Colleen E. Giblin. ``Explanations of the
  endowment effect: an integrative review.'' Trends in cognitive
  sciences 19.6 (2015): 339-348.
\item
  Horowitz, John K., and Kenneth E. McConnell. ``A review of WTA/WTP
  studies.'' Journal of environmental economics and Management 44.3
  (2002): 426-447.
\item
  Bordalo, Pedro, Nicola Gennaioli, and Andrei Shleifer. ``Salience in
  experimental tests of the endowment effect.'' The American Economic
  Review 102.3 (2012): 47-52.
\item
  Price, Curtis R., and Roman M. Sheremeta. ``Endowment origin,
  demographic effects, and individual preferences in contests.'' Journal
  of Economics \& Management Strategy 24.3 (2015): 597-619.
\item
  ``American Psychological Association (2013). Glossary of psychological
  terms''. Apa.org.
\end{enumerate}

\end{document}